\documentclass[prl,twocolumn,superscriptaddress,english]{revtex4-1}
\usepackage[latin9]{inputenc}
\usepackage{amsmath,amssymb,amsthm,amsfonts,mathrsfs}
\usepackage{bm}
\usepackage{CJK}
\usepackage{amstext,bm}
\usepackage{graphicx}
\usepackage{xcolor}
\usepackage{color}
\usepackage{amsmath}
\usepackage{babel}
\usepackage[titletoc]{appendix}
\usepackage{hyperref}
\usepackage{float}
\hypersetup{
     colorlinks = true,
     citecolor  = magenta,  % cite
     linkcolor  = magenta % ref
}

\begin{document}

\title{Nodal Manifolds Bounded by Exceptional Points on Non-Hermitian Honeycomb Lattices and Electrical-Circuit Realizations}% Force line breaks with \\
\author{Kaifa Luo}
\affiliation{School of Physics and Technology, Wuhan University, Wuhan 430072, China}
\author{Jiajin Feng}
\affiliation{School of Physics, Sun Yat-sen University, Guangzhou 510275, China}
\author{Y. X. Zhao}
\email[]{zhaoyx@nju.edu.cn}
\affiliation{National Laboratory of Solid State Microstructures and department of Physics, Nanjing University, Nanjing, 210093, China}
\affiliation{Collaborative Innovation Center of Advanced Microstructures, Nanjing University, Nanjing 210093, China}
\author{Rui Yu}%
\email[]{yurui@whu.edu.cn}
\affiliation{School of Physics and Technology, Wuhan University, Wuhan 430072, China}

%\date{\today}

\begin{abstract}
	Topological semimetals feature a diversity of nodal manifolds including nodal points,
	various nodal lines and surfaces, and recently novel quantum states in non-Hermitian systems
	have been arousing widespread research interests.
	In contrast to Hermitian systems whose bulk nodal points must form closed manifolds,
	it is fascinating to find that for non-Hermitian systems exotic nodal manifolds can be bounded by exceptional points
	in the bulk band structure. Such exceptional points, at which energy bands coalesce with band conservation violated,
	are iconic for non-Hermitian systems.
	In this work, we show that a variety of nodal lines and drumheads with exceptional boundary can be realized
	on 2D and 3D honeycomb lattices through natural and physically feasible non-Hermitian processes.
	The bulk nodal Fermi-arc and drumhead states, although is analogous to, but should be essentially distinguished from the
	surface counterpart of Weyl and nodal-line semimetals, respectively, for which surface nodal-manifold bands
	eventually sink into bulk bands.
	Then we rigorously examine the bulk-boundary correspondence of these exotic states with open boundary condition,
	and find that these exotic bulk states are thereby undermined, showing the essential importance
	of periodic boundary condition for the existence of these exotic states.
	As periodic boundary condition is non-realistic for real materials, we furthermore propose
	a practically feasible electrical-circuit simulation, with non-Hermitian devices implemented by ordinary operational amplifiers, to emulate these extraordinary states.
\end{abstract}

\maketitle

% \tableofcontents
%%========================MAIN TEXT================================

\section{Introduction}
Recently novel quantum states of non-Hermitian systems have been a rapidly expanding field,
acceleratingly attracting attention from the previously unrelated fields,
such as topological phases of quantum matter
\cite{
heiss_exceptional_2004,
QuantumWalk_rudner_topological_PRL_2009,
quantum_wires_diehl_NP_2011,
non_Hermitian_honeycombLattice_PRB_2011,
Heiss_ExceptionalPoint_physics_2012,
review_bardyn_NJP_2013,
yuce_topological_2015,
zeuner_observation_2015,
Tonylee_anomalous_2016,
menke_topological_2017,
ChongYD_leykam_edge_PRL_2017,
Weyl_nonHermitian_gonzalez_PRB_2017,
ChongYD_exceptional_PRB_2017,
Ueda_PRX_2018,
FuLiang_non_hermitian_PRL_2018,
NL_nonHermitian_molina_PRL_2018,
Review_Torres_2018,
Weyl_nonHermitian_Zyuzin_PRB_2018,
Weyl_nonHermitian_FanShanhui_PRB_2018,
ZhangHaiJun_non_hermitian_nodalLine_2018},
many-particle physics
\cite{
FuLiang_disorder_non_hermitian_2017,
FuLiang_interaction_nonhermitian_2018,
EE_non_hermitian_PRB_2018,
FuLiang_shen_quantum_2018},
cold atoms
\cite{
cold_atoms_budich_PRA_2015,
ColdAtom_XuYong_PRL_2017,
cerjan_experimental_2018},
and the traditional field of quantum optics
\cite{
photonic_ruter_observation_NP_2010,
choi_quasieigenstate_2010,
zhen_spawning_2015,
photonic_longhi_EPL_2018,
photonic_review_LuLing_2018,
xing_spontaneous_2017,
harari_topological_2018,
beyna_topological_2018,
bahari_nonreciprocal_2018} with renewed interests.
Maybe the most iconic feature of non-Hermitian physics is the existence of exceptional points~\cite{BOOK_kato_EP_1995} in parameter space,
at which unitary or more general similarity transformations cannot convert the Hamiltonian under consideration into
a completely diagonal form, but optimally into upper-triangular blocks each with equal diagonal entries,
namely a Jordan normal form~\cite{Heiss_ExceptionalPoint_physics_2012}.
Therefore, for a band theory a number of energy bands coalesce at an exceptional point
in the Brillouin zone (BZ), where accordingly energy-band conservation is violated.
On the other hand, the recently enhanced interest in non-Hermitian physics partially evolved from topological phases
of quantum matter, where topological semimetals as a central topic feature nodal manifolds in the BZ including
degenerate nodal lines~\cite{
mikitik_band_contact_2006,
burkov_topological_2011,
lu_weyl_2013,
zhao2013topological,
chiu_classification_2014,
zhao2016unified,
yang_dirac_2014,
heikkila_nexus_2015,
fangchen_topological_2015,
xie_new_2015,
weng_topological_2015,
mullen_line_2015,
TNL_Shanhui_Zahid_PRB_2016}
and surfaces~\cite{
Topo_surfaceState_Mele_2013,
Topo_surfaceState_Shanhui_2017,
Topo_surfaceState_Murakami_NC_2017,
Topo_surfaceState_Sigrist_PRB_2017,
Topo_surfaceState_Shengyuan_2018,
Topo_surfaceState_Shengyuan_2018,
Topo_surfaceState_Sergej_2018}.
Due to band conservation of Hermitian theory, such nodal manifolds are always closed and accordingly have no boundary.
Now considering nodal manifolds in non-Hermitian systems, one may expect an exotic quantum state solely for non-Hermitian system,
namely, that nodal manifold can terminate on a boundary consisting of exceptional points
\cite{
twisted_Fermi_ribbons_carlstrom_PRA_2018,
HuJiangPing_nonHermitian_2018,
2D_EL_2018,
Fermi_ribbon_disorder_driven_2018},
and indeed recently the bulk Fermi arc, which is an open nodal ended with two exceptional points
\cite{
FuLiang_disorder_non_hermitian_2017,
FuLiang_interaction_nonhermitian_2018},
have been realized in non-Hermitian photonic crystals with much attention attracted
\cite{FuLiang_bulk_fermi_arc_exp_science_2018}.
In this article we show that a variety of open nodal manifolds with exceptional boundaries,
including various Fermi arcs and particularly drumheads, namely open surfaces, can be realized in the bulk band structures
of 2D and 3D honeycomb lattices through natural and physically feasible non-Hermitian processes.
Our models are quite simple with only nearest-neighbor hoppings included,  %and next-nearest-neighbor
and may be understood as non-Hermitian theories of graphene and graphite.
% \cite{non_Hermitian_graphene_PRB_2011}.

It is also interesting to compare the bulk nodal Fermi arcs and drumhead states with the boundary Fermi arcs and drumhead states
of Weyl and nodal-line semimetals, respectively. Although for both cases they are open manifolds,
Hermitian systems preserve band number and therefore the open manifolds of boundary band structure necessarily sink into
and connect with the bulk energy bands. But maybe more profoundly the boundary of a Hermitian system is not an independent
system, and in this sense it might bear certain connections with non-Hermitian physics that is essentially devoted
to open systems.

Recently it is noticed that the physical property of non-Hermitian systems can be radically dependent
on boundary conditions
\cite{
Tonylee_anomalous_2016,
boundary_xiong_why_2018,
Torres_1D_PRB_2018,
SSH_Simon_PRB_2018,
SSH_Yuce_PRA_2018,
SSH_ChenShu_geometrical_PRA_2018,
WangZhong_1D_nonHermitian_PRL_2018}.
For instance, the spectrum of the Su-Schrieffer-Heeger (SSH) model with small anti-Hermitian nearest-neighbor hoppings is complex under
periodic boundary conditions, but is purely real under open boundary conditions.
As the representation by the BZ actually corresponds to periodic boundary conditions, we proceed to study the bulk-boundary
correspondence of our non-Hermitian honeycomb-lattice models with open boundary conditions, and find that the non-Hermitian
tight-binding models are equivalent to Hermitian ones by similarity transformations.
This shows that the exotic quantum states of nodal manifolds bounded by exceptional points can be undermined by open boundary
conditions, and periodic boundary conditions are therefore essential for their existence.
To circumvent the dilemma that periodic boundary conditions are not realistic for ordinary physical systems, such as real materials, photonic crystals, phononic crystals and cold atoms, and also inspired by
that novel band theory has broader applications beyond electronic systems, we present a simulation of the non-Hermitian
tight-binding models on honeycomb lattices through faithfully designating electrical circuits,
for which periodic boundary conditions are obviously realizable
\cite{
Topological_Circuit_PRX_2015,
Topological_Circuit_PRL_2015,
Topolectrical_Circuits_Weyl_Ronny_CoomP_2018}.
Particularly non-Hermitian devices, emulating non-Hermitian terms, can be easily implemented via a standard application
of a common operational amplifier in a voltage follower configuration.
Furthermore, the feasibility of the particular design is ensured by the fact that each unit cell only
consists of a few capacitors, inductors and operational amplifiers.

The article is organized as follows.
Section~\ref{sec:lattice} briefs the 2D and 3D non-Hermitian lattice structures, which are used in the
following discussions.
In Sec.~\ref{sec:2D_case} we show the exceptional points and bulk Fermi-arc states terminated at exceptional end-points in the 2D structure.
In Sec.~\ref{sec:3D_case} we investigate the exceptional lines and bulk drumheads states with exceptional edges in the 3D structure.
In Sec.~\ref{sec:OBC_case} we compare the band dispersions between periodic boundary conditions and open boundary conditions, and show that the periodic boundary conditions are essential for the above bulk states.
Finally Sec.~\ref{sec:circuit} presents the designed non-Hermitian electrical-circuit lattices, which are easy to achieve periodic boundary conditions in 2D and 3D cases, and can realize bulk quantum states of nodal manifolds bounded by exceptional points.

\section{2D and 3D non-hermitian honeycomb lattices\label{sec:lattice}}
\begin{figure}[h]
			\centering
			\includegraphics[width=0.95\linewidth]{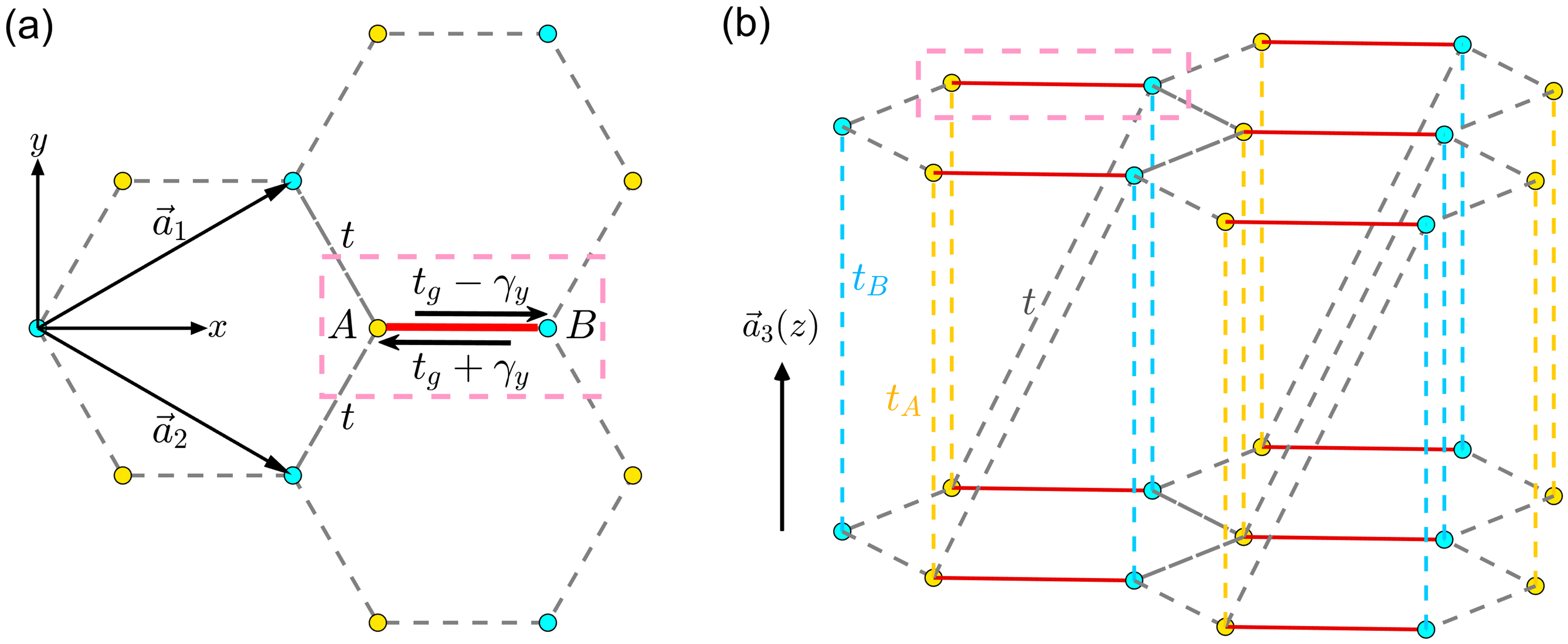}
			\caption{\label{fig:lattice_2D_3D} (a) 2D and (b) 3D honeycomb lattice.
			The dashed pink box indicates the unit cell.
			The Hopping parameters $t_g\pm \gamma_y$ inside a unit cell are asymmetric, leading to the non-Hermitian term.
			The interactions between unit cells on the 2D plane are set as $t$.
			The inter-layer couplings are set as $t_A$, $t_B$ and $t$ between A-A, B-B and A-B sites, respectively.
			}
			\end{figure}

Honeycomb lattice plays an important role in constructing models of novel topological quantum states.
For instance, electrons on 2D honeycomb lattices may have the Dirac-type energy dispersions,
which have aroused tremendous research interests for topological phases.
From the topological point of view, the massless Dirac point usually corresponds to criticality of phase transition between two topologically distinct phases.
Notably both
the quantum spin Hall states~\cite{kane_QSH_2005} and
the quantum anomalous Hall state~\cite{haldane_model_1988}
were first proposed in the 2D honeycomb lattice as pioneering models of topological insulators, which is in retrospect based on the Dirac criticality.
It is also a good starting point to look for nodal-line and Weyl semimetal semimetal phases on 3D
honeycomb lattices formed by stacking 2D honeycomb lattices along the vertical dimension~\cite{luo_topological_2018}.
As aforementioned honeycomb-lattice models are all Hermitian, the dissipative (gain/loss) and nonreciprocal effects are not taken into consideration.
In this work, we demonstrate that, in the non-Hermitian regime, honeycomb lattices are a cornerstone as well for seeking novel quantum states, which essentially depend on non-Hermiticity.
As shown in Fig.~\ref{fig:lattice_2D_3D}, both 2D and 3D honeycomb lattices consist of sublattices A and B, and the unit cell is indicated in the pink-dashed box.
We assume the hopping processes within each unit cell are asymmetric for sublattices A and B, resulting in the non-Hermitian terms,
while the hoppings between unit cells, which are are symmetric, lead to the corresponding Hermitian terms.

%% fig---1
%% fig---1

\section{Exceptional points and bulk Fermi-arc states in 2D non-Hermitian honeycomb lattice \label{sec:2D_case}}

\begin{figure}[h]
			\centering
			\includegraphics[width=0.95\linewidth]{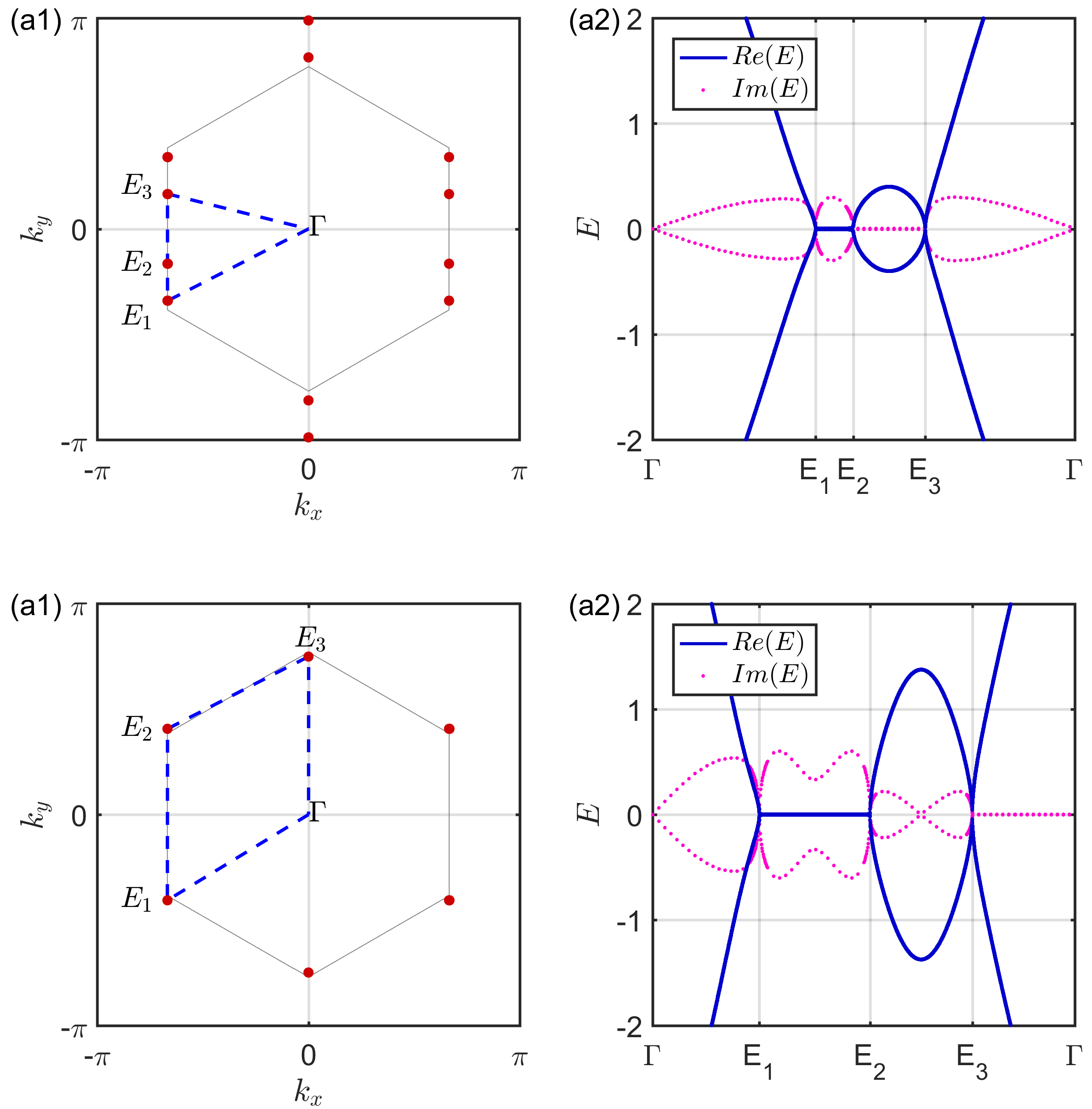}
			\caption{\label{fig:2D_EP} The solutions of Eq.~(\ref{eq:2D_exceptional_points_2}) with (a) 4 and (b) 2 exceptional points (red dots) in the first Brillouin zone. The parameters are setting as $t=1,t_g=1.5$, and (a) $\gamma_y=0.3$ (b) $\gamma_y=0.6$. The real and imaginary part of the band dispersions along the line through the exceptional points are showing in (a2) and (b2).}
			\end{figure}

% \subsection{Exceptional points}
We begin with the 2D case, for which the tight-binding Hamiltonian is written as
	\begin{equation}
	H(\bm{k})=d_x(\bm{k})\sigma_x+\left(d_y(\bm{k})+i\gamma_y\right)\sigma_y,
	\label{eq:HK_2D_model}
	\end{equation}
where
$d_x=t_g+t(\cos\bm{k}\cdot\bm{a}_1+\cos\bm{k}\cdot\bm{a}_2)$,
$d_y=    t(\sin\bm{k}\cdot\bm{a}_1+\sin\bm{k}\cdot\bm{a}_2)$,
$\bm{a}_{1,2}=a\left(3/2,\pm\sqrt{3}/2\right)$
are the lattice vectors and we set the atom-atom distance $a=1$ hereafter.
$t_g$, $t$ and $\gamma_y$ are hopping parameters as indicated in Fig.~\ref{fig:lattice_2D_3D}(a).
$\sigma_{x,y,z}$ are the Pauli matrices while the term involving $\sigma_z$ vanishes
under the assumption of chiral symmetry of the system.
The energy dispersions are then calculated as
	\begin{equation}\label{eq:EK_2D}
	E_{\pm}(\bm{k})=\pm\sqrt{d^2_x(\bm{k})+d^2_y(\bm{k})-\gamma_y^2+2id_y(\bm{k})\gamma_y},
	\end{equation}
which is generally complex for nonzero $\gamma_y$.
The exceptional point appears if two bands coalesce, leading to
	\begin{equation}\label{eq:2D_exceptional_points_1}
	d_y(\bm{k})=0 ~~and ~~d_x(\bm{k})=\pm \gamma_y,
	\end{equation}
and can be combined into a single complex equation
	\begin{equation}\label{eq:2D_exceptional_points_2}
	d_x+id_y=t_g+t(e^{i\bm{k}\cdot\bm{a}_1}+e^{i\bm{k}\cdot\bm{a}_2})=\pm \gamma_y.
	\end{equation}
By tuning the parameters, we get different numbers of solutions for Eq.~\eqref{eq:2D_exceptional_points_2},
i.e., different number of exceptional points in the BZ:
1) $\mathrm{max}[|(t_g\pm\gamma_y)/t|]<2$, there are four exceptional points in the first BZ (Fig.~\ref{fig:2D_EP}(a1)).
2) $\mathrm{min}[|(t_g\pm\gamma_y)/t|]  < 2 < \mathrm{max}[(t_g\pm\gamma_y)/t]$, two exceptional points appear (Fig.~\ref{fig:2D_EP}(b1)).
3) No exceptional point exists when $\mathrm{min}[|(t_g\pm\gamma_y)/t|]>2$.
The band dispersions through the exceptional points are shown in Fig.~\ref{fig:2D_EP}(a2, b2).

%% fig---2

\begin{figure}[h]
	    	\centering
	    	\includegraphics[width=0.95\linewidth]{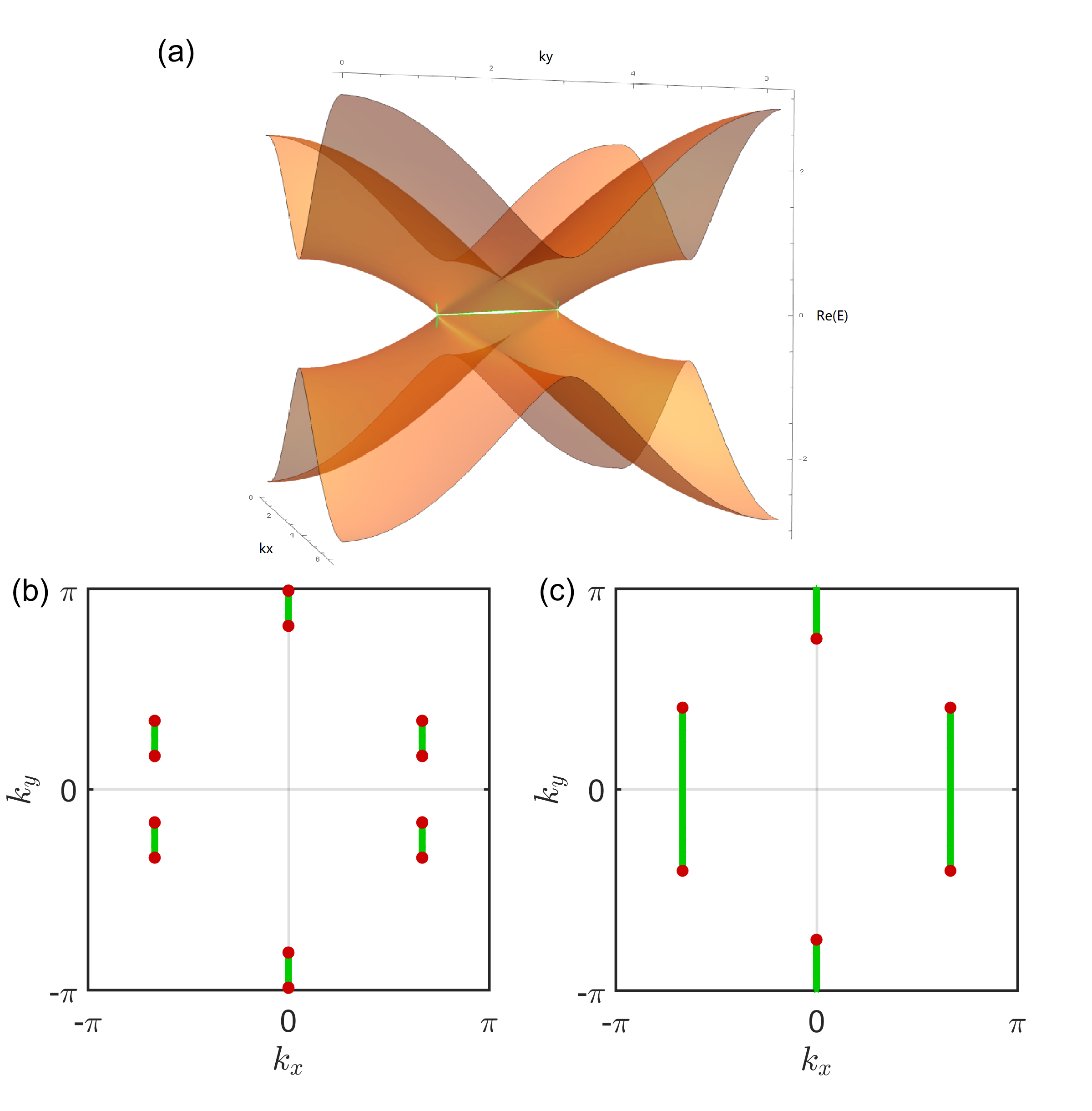}
	    	\caption{\label{fig:2D_BFA} (a) The real part of the
	    	spectrum, which are degenerate along a line,
         form the so-called bulk Fermi-arc.
	    	(b) and (c) are the configurations of the bulk Fermi-arc with the parameters
	    	the same as Fig.~\ref{fig:2D_EP}(a) and (b), respectively.
	    	Each bulk Fermi-arc (green lines) is ended with two exceptional points (red points).}
	    \end{figure}

% \subsection{Bulk Fermi-arc states}
Quite different from the Hermitian systems, a pair of exceptional points in the complex spectrum of non-Hermitian Hamiltonian will lead to an open-ended bulk states, i.e., the so-called bulk Fermi-arc
\cite{
FuLiang_disorder_non_hermitian_2017,
FuLiang_interaction_nonhermitian_2018,
FuLiang_bulk_fermi_arc_exp_science_2018}.
As schematically plotted in Fig.~\ref{fig:2D_BFA}(a), the bulk Fermi-arc, degenerate with real part of the eigenvalues while non-degenerate with the imaginary part, links a pair of exceptional points.
The number of bulk Fermi-arcs and exceptional points can be tuned by the parameter $\gamma_y$ as discussed above.
Below, we demonstrate the existence of bulk Fermi-arc in our model of Eq.~\eqref{eq:HK_2D_model}.
From the dispersion expression Eq.~\eqref{eq:EK_2D}, it is easy to find that if the following expressions
	\begin{equation}\label{eq:2D_bulk_fermi_arc}
	\mathrm{Re}\left(E_{\pm}^2\right)<0~{\rm and}~\mathrm{Im}\left(E_{\pm}^2\right)=0
	\end{equation}
are satisfied, the real parts of the dispersions are degenerate while the imaginary parts are non-degenerate, which are the solutions for the bulk Fermi-arc.
Substituting Eq.~(\ref{eq:EK_2D}) into Eq.~(\ref{eq:2D_bulk_fermi_arc}), we obtain
	\begin{equation}\label{eq:2D_bulk_fermi_arc2}
	d_y(\bm k)=0~{\rm and}~d_x^2(\bm k)<\gamma_y^2.
	\end{equation}
Comparing Eq.~(\ref{eq:2D_exceptional_points_1}) with Eq.~(\ref{eq:2D_bulk_fermi_arc2}),
it is obvious that the exceptional points are the boundaries of bulk Fermi-arcs.
Solving Eq.~(\ref{eq:2D_bulk_fermi_arc2}), one obtains the explicit ranges of bulk Fermi-arc in the BZ, which read
1) $k_x=4n\pi/3$, $|t_g + 2 t \cos(\sqrt{3}k_y/2)|<|\gamma_y|$;
2) $k_x=(4n+2)\pi/3$, $|t_g - 2 t \cos(\sqrt{3}k_y/2)|<|\gamma_y|$; and
3) $k_y=(2n+1)\pi/\sqrt{3}$, $|t_g|<|\gamma_y|$.
The configurations of the bulk Fermi arcs for $\gamma_y=0.3$ and $0.6$ are shown in Fig.~\ref{fig:2D_BFA}(b, c).

%% fig---3

%%non-Hermitian 3D Honeycomb Lattice
\section{Exceptional Lines, Bulk drumhead states in 3D non-Hermitian Honeycomb Lattices\label{sec:3D_case}}
Inspired by the existence of bulk Fermi-arc terminated at the exceptional points in 2D BZ,
for the 3D honeycomb lattice, due to the increasing of spatial dimensionality,
we expect to obtain lines of exceptional points and drumhead-like bulk states bounded by this exceptional lines.
The tight-binding Hamiltonian for the lattice model given in Fig.~\ref{fig:lattice_2D_3D}(b) is
	\begin{equation}\label{eq:3D_HK_OBC}
	H(\bm{k})=d_x(\bm{k})\sigma_x+\left(d_y(\bm{k})+i\gamma_y\right)\sigma_y+d_z(\bm{k})\sigma_z,
	\end{equation}
where
$d_x(\bm{k})=t_g+t\sum_{j=1}^{3}\cos\bm{k}\cdot\bm{a}_j$,
$d_y(\bm{k})=t\sum_{j=1}^{3}\sin\bm{k}\cdot\bm{a}_j$,
$d_z(\bm{k})= t_{AB}\cos\bm{k}\cdot \bm{a}_3+\mu_{AB}$,
$t_{AB}\equiv t_A-t_B$, and
$\mu_{AB}\equiv(\mu_A-\mu_B)/2$
is introduced to indicate the difference of the on-site energies between sublattices $A$ and $B$.
In the following, we investigate this model for two cases.

\begin{figure}[b]
	    	\centering
	    	\includegraphics[width=0.95\linewidth]{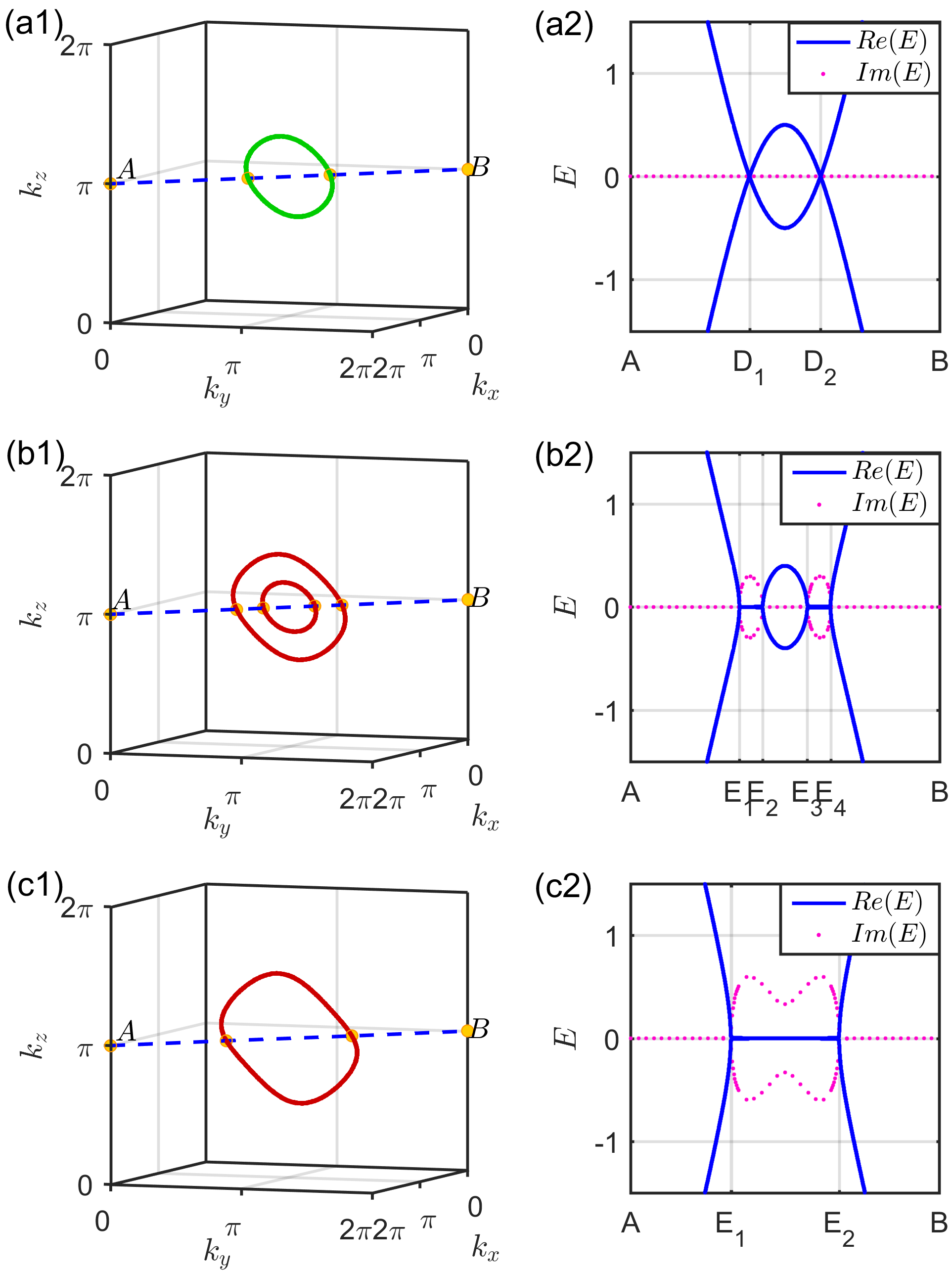}
	    	\caption{\label{fig:3D_Exceptional_Lines_dz0}
	    	 Fixing $t_g=2.5,~t=1$ and tuning the value of parameter $\gamma_y$, one obtain
	    	 (a1) Ring shape bands degenerate points for $\gamma_y=0$. In this case the system is Hermitian, the energy of bands are real as show in (b2).
	    	 (b1) The exceptional points form two rings in the BZ for $\gamma_y=0.3$ which satisfy $max[|(t_g\pm\gamma_y)/t|]<3$. The complex bands is square-root nearby the exceptional points as shown in (b2).
	    	 (c1) and (c2) For $\gamma_y=0.6$, which satisfy $min[|(t_g\pm\gamma_y)/t|] < 3  < max[|(t_g\pm\gamma_y)/t|]$, only one exceptional ring left.}
	    \end{figure}

% \subsection{Exceptional lines and bulk drumhead states in the case of $d_z=0$}
We first consider a simplified case, where $d_z=0$, i.e., setting $t_{AB}=\mu_{AB}=0$.
Then the exceptional points are the solutions of the following equations
	\begin{equation}\label{eq:3D_Exceptional_Lines_dz0}
	d_x(\bm{k})+id_y(\bm{k})=t_g+h(\bm{k})=\pm\gamma_y
	\end{equation}
where
$h(\bm{k}) \equiv t \sum_{j=1}^{3}e^{i\bm{k}\cdot\bm{a}_j}$.
If $\gamma_y=0$, the solutions of Eq.~\eqref{eq:3D_Exceptional_Lines_dz0} form a nodal-ring in momentum space (shown in Fig.~\ref{fig:3D_Exceptional_Lines_dz0}(a)) for appropriate values of $t_g$ and $t$ as discussed in \cite{luo_topological_2018}.
For a nonzero $\gamma_y$, there are three types of solutions.
1) $\mathrm{max}[|(t_g\pm\gamma_y)/t|]<3$, the solutions of Eq.~\eqref{eq:3D_Exceptional_Lines_dz0} form two exceptional rings in the 3D BZ, as shown in Fig.~\ref{fig:3D_Exceptional_Lines_dz0}(b).
2) $\mathrm{min}[|(t_g\pm\gamma_y)/t|] < 3  < \mathrm{max}[|(t_g\pm\gamma_y)/t|]$, a single exceptional ring exists, as shown in Fig.~\ref{fig:3D_Exceptional_Lines_dz0}(c).
3) $\mathrm{min}[|(t_g\pm\gamma_y)/t|]>3$, no exceptional point solutions exist for Eq.~(\ref{eq:3D_Exceptional_Lines_dz0}).
%% fig---4

\begin{figure}[h]
	    	\centering
	    	\includegraphics[width=0.95\linewidth]{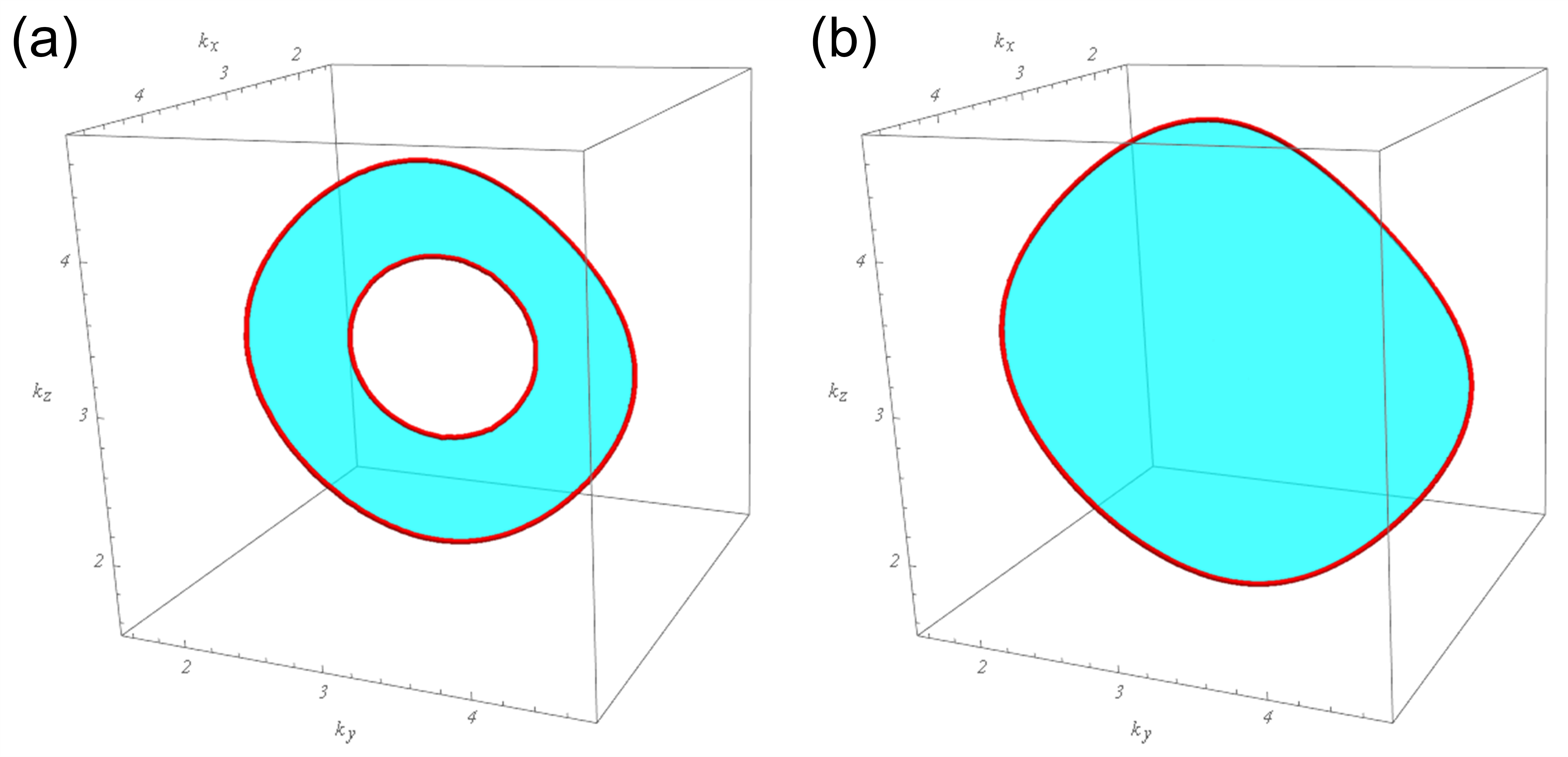}
	    	\caption{The bulk drumhead states are the states with eigenvalues degenerate for the real parts but splitted for the imaginary parts. The bulk drumhead states (cyan color surface) bounded by exceptional rings (red color curves) for (a) $\gamma_y=0.3$ with a hole and (b) $\gamma_y=0.6$ without hole.
	        }\label{fig:3D_Bulk_Nodal_Surface_dz_0}
	    \end{figure}

Substituting Eq.~(\ref{eq:3D_HK_OBC}) into Eq.~(\ref{eq:2D_bulk_fermi_arc}), we obtain
	\begin{equation}\label{eq:3D_nodal_surface_dz0}
	\left(t_g+h(\bm{k})\right)^2<\gamma_y^2.
	\end{equation}
The solutions of Eq.~\eqref{eq:3D_nodal_surface_dz0} determine the range of the desired drumhead states.
Comparing Eq.~(\ref{eq:3D_Exceptional_Lines_dz0}) with Eq.~(\ref{eq:3D_nodal_surface_dz0}), we obtain that the exceptional rings are the boundary of the bulk drumhead states.
The configuration of bulk drumhead states is dependent on the parameters as discussed
below Eq.~(\ref{eq:3D_Exceptional_Lines_dz0}).
By numerically solving Eq.~(\ref{eq:3D_nodal_surface_dz0}), we find two types of bulk drumhead states.
The first type is a drumhead with a hole, bounded by two exceptional lines
(Fig.~\ref{fig:3D_Bulk_Nodal_Surface_dz_0}(a)).
The second type is a whole drumhead bounded by one exceptional line (Fig.~\ref{fig:3D_Bulk_Nodal_Surface_dz_0}(b)).
These bulk drumhead states are essentially different from the drumhead surface states in the Hermitian nodal-line semimetals.
For the latter, the degenerate points form
nodal-rings in the 3D bulk BZ, and due to the bulk-boundary correspondence, lead to the drumhead boundary states on the 2D surface BZ, whose edges eventually sink into and connect with the bulk nodal-line states.
While for the 3D non-Hermitian system,
the drumhead states are bulk states bounded by the exceptional lines, with eigenvalue degenerate for the real part but splitted for the imaginary part.

%% fig---5

%%Exceptional Points
% \subsection{Exceptional lines and bulk \yu{drumheads} states in the case of $d_z\ne 0$}

\begin{figure}[]
	        \centering
	        \includegraphics[width=0.75\linewidth]{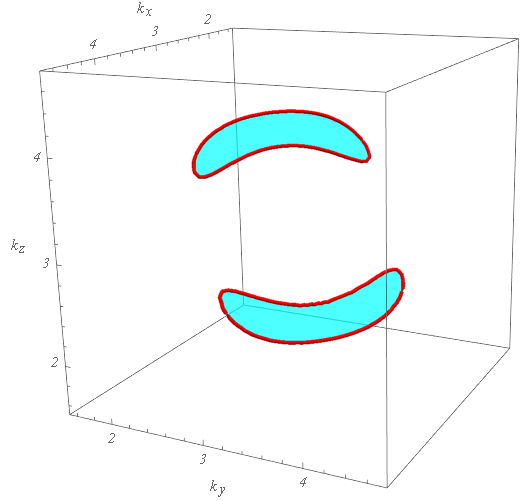}
	        \caption{\label{fig:3D_EL_and_BNS_dz_1}
	        The exceptional lines (red color curves) and the bulk drumhead states (cyan color surface) for Eq.~(\ref{eq:3D_HK_OBC}) with
	        parameters $t_g=2.5$, $t=t_{AB}=1$, $\mu_{AB}=0.7$ and $\gamma_y=0.2$.
	        }
	    \end{figure}

Now we discuss the more general case with $d_z\ne 0$ in Eq.~(\ref{eq:3D_HK_OBC}).
For Hermitian system with $\gamma_y=0$, Weyl points can be realized in this 3D honeycomb
lattice if $d_x(\bm k)=d_y(\bm k)=d_z(\bm k)=0$ are satisfied~\cite{luo_topological_2018}.
For non-Hermitian system with $\gamma_y \ne 0$, the configuration of drumhead states with exceptional edges is enriched compared to the Hermitian case and the non-Hermitian case of $d_z=0$.
In parallel to the discussions in previous sections,
the exceptional points and the bulk drumhead states are determined by the following equation and in-equation respectively.
	\begin{equation}
	d_y(\bm{k})=0,~d_x^2(\bm{k})+d_z^2(\bm{k})=\gamma_y^2,\label{eq:dz_ne_0_exceptional points}
	\end{equation}
	\begin{equation}
	d_y(\bm{k})=0,~d_x^2(\bm{k})+d_z^2(\bm{k})<\gamma_y^2.\label{eq:dz_ne_0_Real}
	\end{equation}
Setting $\mu_{AB}=0.7$, $\gamma_y=0.2$ and $t_{AB}=1$, we obtain
two drumhead states bounded by two exceptional rings as shown in Fig.~\ref{fig:3D_EL_and_BNS_dz_1}.
% \yu{
% Changing these parameters, the exceptional points can also form two closed rings or two unclosed lines,
% which is discussed in the supplementary materials.
% }
 %\yu{ add the results in SM.}
%% fig---6

\section{The Bulk-Boundary Correspondence\label{sec:OBC_case}}
In the previous two sections, we discussed the band structures of the bulk states, where
the periodic boundary conditions were actually implicitly presumed for the Fourier transforms can be applied to produce the BZ.
For non-Hermitian system, the bulk energy spectra may change dramatically with
open boundary conditions for non-Hermitian systems, which is in sharp contrast to Hermitian ones.
In this section, we take 2D non-Hermitian honeycomb lattice as an example to show how the band structures change form periodic boundary conditions to open boundary conditions, and discuss the bulk-boundary correspondence.
The derivation details are given in Appendix~\ref{appendix:openBC_2D}, and the results for the 3D case are given in Appendix~\ref{appendix:openBC_3D}.

\begin{figure}[t]
	        \centering
	        \includegraphics[width=0.95\linewidth]{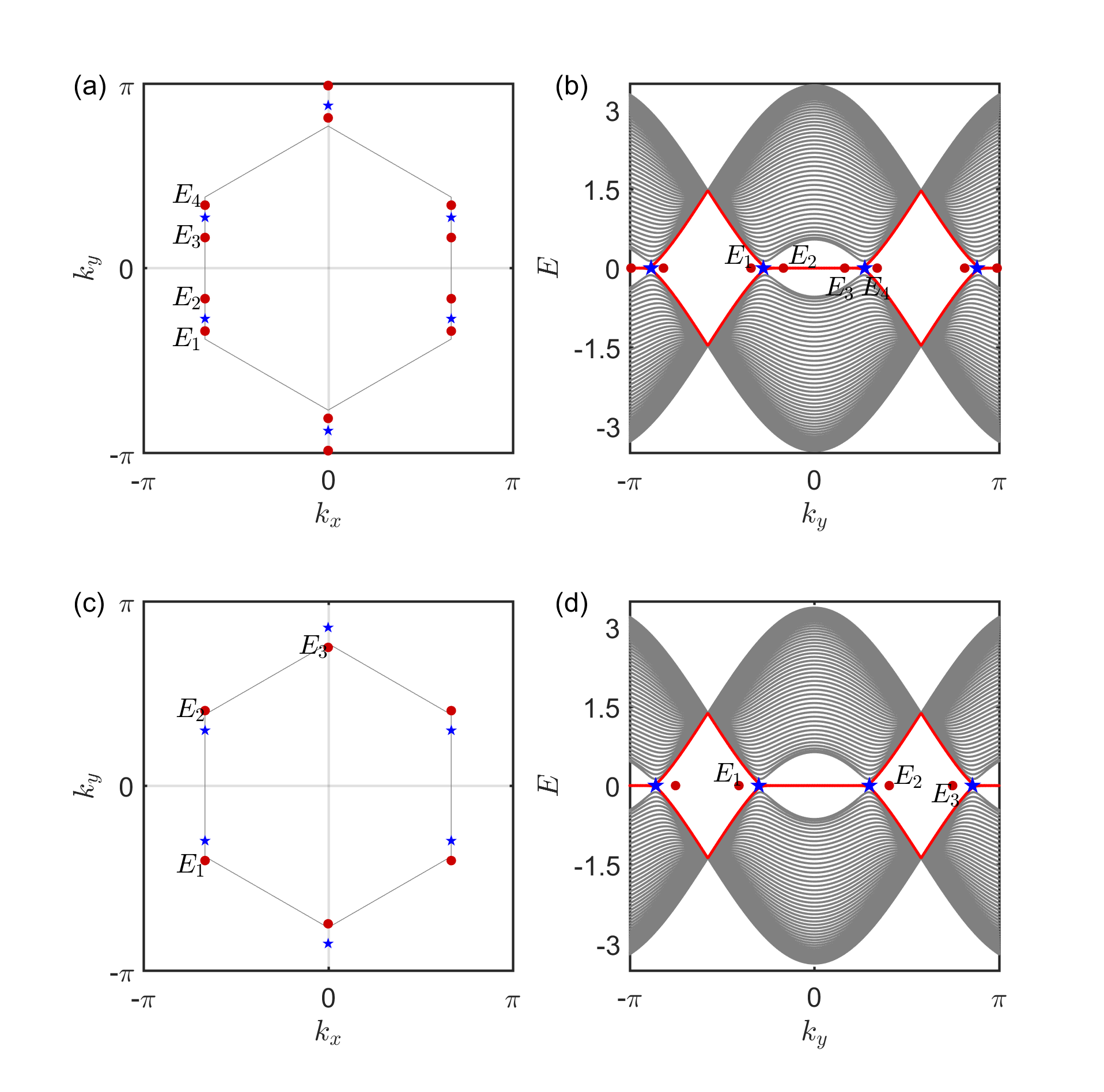}
	        \caption{\label{fig:2D_edge_state} Exceptional points (red points) for Eq.~(\ref{eq:2D_exceptional_points_2}) and Nodal points (blue stars) for Eq.~(\ref{eq:2D_Hermitian_Hk}) with parameter (a, b) $\gamma_y=0.3$ and (c, d) $\gamma_y=0.6$.
	        The edge states (red lines) appear and connect a pair of nodal points for the non-Hermitian 2D honeycomb lattice with zigzag edges.}
	    \end{figure}

The band dispersions for a strip of 2D honeycomb lattice with zigzag edge in the $x$ direction
are shown in Fig.~\ref{fig:2D_edge_state}.
It is observed that the band-crossing points (blue stars) do not correspond to the exceptional points (red dots),
and the number of gap-closing points can be different from that of exceptional points. Significantly the Fermi-arc states connecting exceptional points $E_1$-$E_2$ and $E_3$-$E_4$ (red dots) existing in the periodic boundary conditions
disappear for the strip structure with open boundary conditions.
However, the edge states connecting the band-crossing points (blue stars) now present.

%% fig---7

The above results indicate that the proposed exceptional points
and the bulk Fermi-arc states can only exist in a system with periodic boundary conditions.
This requirement clearly brings difficulty to realize these states experimentally, for the periodic boundary conditions are not easy, if not impossible, to implement in commonly used experimental systems, such as real materials, photonic crystals, phononic crystals, and cold atoms.

To solve this problem, we propose to simulate these states in electrical-circuit lattices, for
which the periodic boundary conditions are quite easy to be implemented if we connect the head with the tail,
showing a significant advantage compared with other realization scenarios.
In the following section, we detail how to design the non-Hermitian honeycomb lattice
to realize the novel states discussed above.

\section{Non-Hermitian electrical circuit lattice\label{sec:circuit}}

Recently, there has been growing interest in realizing
topological phases by electrical circuits, including
the time-reversal-invariant topological insulators
\cite{
Topological_Circuit_PRX_2015,
Topological_Circuit_PRL_2015,
HuXiao_generating_2018},
3D Weyl semimetals
\cite{
	Topolectrical_Circuits_Weyl_Ronny_CoomP_2018,
	luo_topological_2018,
	circuit_Weyl_probing_2018},
1D topological insulators
\cite{Topolectrical_Circuits_ssh_PRB_2018} and
the higher-order topological insulators
\cite{
	Topolectrical_Circuits_Ronny_NP_cornermodes_2018,
	Topolectrical_Circuits_Ezawa_1_2018,
	Topolectrical_Circuits_Ezawa_2_2018}.
In this section, we construct a 2D electrical-circuit lattice, consisting of capacitors, inductors and operational amplifiers,
as an experimental setup to realize the bulk Fermi-arc states bounded by exceptional points discussed in Sec.~\ref{sec:2D_case}.
The 3D electrical-circuit lattice for the bulk drumhead states with exceptional edge states discussed in Sec.~\ref{sec:3D_case} can be designated in a similar way.

% %% fig---8
\begin{figure}[t]
	\centering
	\includegraphics[width=0.95\linewidth]{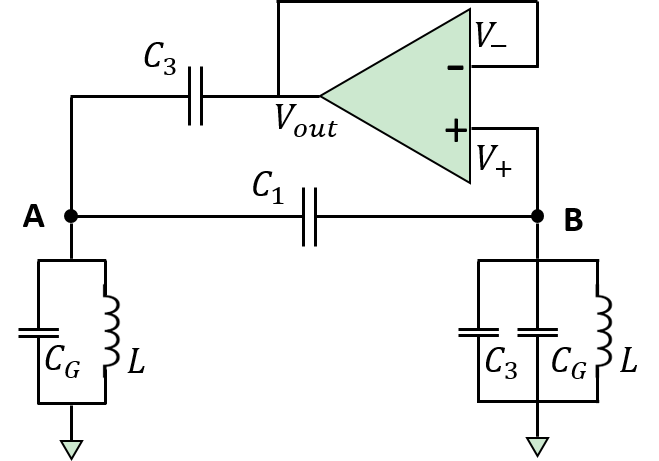}
	\caption{(a) The elementary circuit cell that gives the non-Hermitian effect. The operational amplifier's inputs consist of
		a non-inverting input $(+)$ with voltage $V_{+}$ and
		an    inverting input $(-)$ with voltage $V_{-}$.
		The output voltage of the operational amplifier is denoted as $V_{out}$.
		Connecting the inverting input $(-)$ and the output, the operational amplifier
		is used as voltage follower, which gives $V_{+}=V_{out}$ but no current flows into the non-inverting input $(+)$.
		$C_1$ and $C_3$ are capacitors.
		The nodes A and B are connected to ground through parallel connected capacitor and inductor
		$C_G$, $C_3$, $L$.}\label{fig:circuit_nonHermitian_part}
\end{figure}

\begin{figure}[h]
	\centering
	\includegraphics[width=0.95\linewidth]{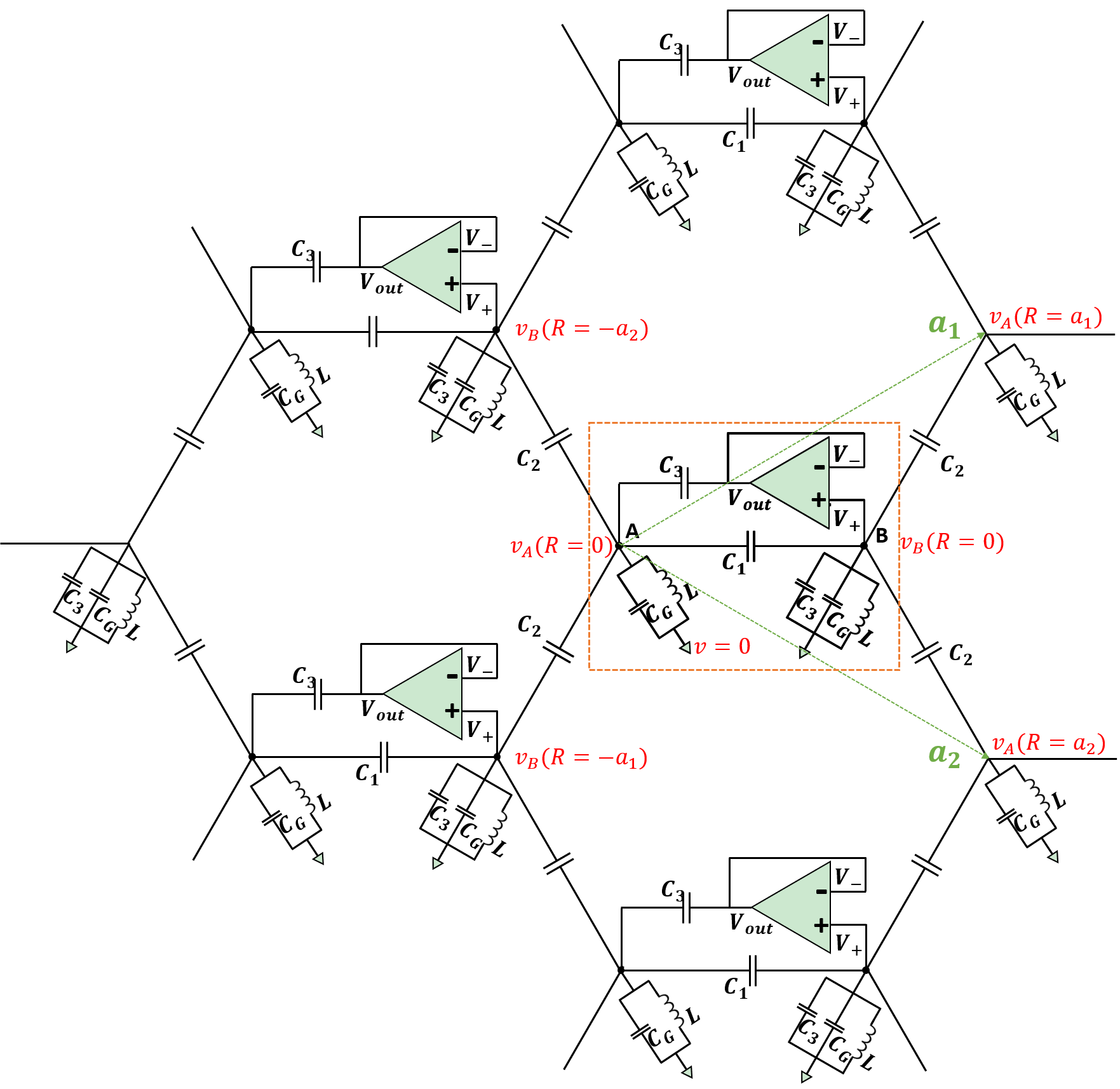}
	\caption{A 2D circuit lattice consist of operational amplifiers and capacitors.
		The dashed pink box indicates the elementary non-Hermitian unit cells.
		The capacitors $C_2$ connect the unit cell, forming a honeycomb-type lattice.}\label{fig:2D_nonHermitian_CircuitLattice}
\end{figure}

We first elaborate how to construct the elementary circuit cell, which corresponds to the non-Hermitian term, as shown in Fig.~\ref{fig:circuit_nonHermitian_part}. The key idea is to utilize operational amplifiers, which are standard components in electrical circuits, to emulate gain and loss, the characteristics of non-hermiticity. Hence, let us begin with some basics of operational amplifier.
The differential inputs of the operational amplifier are characterized by a
non-inverting input $(+)$ with voltage $V_{+}$ and
an inverting input $(-)$ with voltage $V_{-}$.
Ideally the operational amplifier amplifies the difference in voltage between the two inputs.
The output voltage of the operational amplifier $V_{out}$ is given by the equation
$V_{out}=A(V_+-V_-)$, where $A$ is the open-loop gain of the amplifier that is very high for an ideal amplifier.
Connecting the inverting input $(-)$ and the output, leading to $V_{-}=V_{out}$, the amplifier is used as a voltage follower, for that $V_{out}=A(V_+-V_{out}) \Rightarrow V_{out}=\frac{A}{A+1}V_{+}  \approx  V_{+}$.
For an ideal operational amplifier, there is no voltage across its inputs.
Therefore the input terminals $V_{+}$ and $V_{-}$ behave like a short circuit.
But this kind of short is virtual, different from a real one, and draws no current because of
the infinite impedance between the two inputs.
With these properties and according to Kirchhoff's current law, we get the following equations
\begin{equation}
I_{A}=j\omega (C_{3}+C_{1})(v_{B}-v_{A})+j\omega C_{G}(0-v_{A})+\frac{1}{j\omega L}(0-v_{A}),\label{eq:IA}
\end{equation}
\begin{equation}
I_{B}=j\omega C_{1}(v_{A}-v_{B})+j\omega (C_{3}+C_{G})(0-v_{B})+\frac{1}{j\omega L}(0-v_{B}),\label{eq:IB}
\end{equation}
where $\omega$ is the frequency of voltage and $j\equiv\sqrt{-1}$. Considering the current conservation, namely, that the summation of the inflow and outflow
currents at every node is zero, these equations can be simplified, and then recast into the matrix form,
\begin{equation}
\left[\begin{array}{cc}
(C_{1}+C_{3}+C_{G}) & -C_{1}-C_{3}\\
-C_{1} & (C_{1}+C_{3}+C_{G})
\end{array}\right]\left[\begin{array}{c}
v_{A}\\
v_{B}
\end{array}\right]=\frac{1}{\omega^{2}L}\left[\begin{array}{c}
v_{A}\\
v_{B}
\end{array}\right].\label{eq:Kirchhoff_EQ}
\end{equation}
The two-by-two matrix on the right hand of Eq.~\eqref{eq:Kirchhoff_EQ} is clearly non-Hermitian because it is real but not symmetric. Hence, a non-Hermitian device has been constructed by using conventional operational amplifiers, and
% Form this Hamiltonian, one can calculate the eigenvalues $\omega$ and the right-eigenvector $[v_B, v_B]^T$.
repeating this elementary non-Hermitian cell, we can build the 2D non-Hermitian honeycomb lattices. Consequently, the desired electrical-circuit lattice can be constructed as illustrated in  Fig.~\ref{fig:2D_nonHermitian_CircuitLattice}, which is made of the elementary non-Hermitian cells and capacitors $C_2$.
%% fig---9

Now periodic boundary conditions can be readily imposed on the 2D electric-circuit lattice by accordingly connecting components on the left (upper) edge to those on the right (lower) edge.  And the Fourier transforms can be performed, so that the Kirchhoff equations  can be expressed into an eigenvalue-like equation for the stationary systems
\begin{equation}
H(\bm{k})V=\frac{1}{\omega^{2}L}V\label{eq:HV=eV},
\end{equation}
where
\begin{equation}
H(\bm{k})=C_s\sigma_0+d_{x}(\bm{k})\sigma_{x}+(d_{y}(\bm k)+i\gamma_{y})\sigma_{y},\label{eq:Hk_sigma}
\end{equation}
and
\begin{equation}
\begin{aligned}
d_{x}(\bm{k})&=-(C_{1}+\frac{C_{3}}{2})-C_{2}(\cos\bm{k}\cdot\bm{a}_{1}+\cos\bm{k}\cdot\bm{a}_{2}),\\
d_{y}(\bm{k})&=-C_{2}(\sin\bm{k}\cdot\bm{a}_{1}+\sin\bm{k}\cdot\bm{a}_{2}),\\
\gamma_{y}&=-\frac{C_{3}}{2}.\label{eq:d1d2gamma}
\end{aligned}
\end{equation}
Here, $C_s=C_{1}+2C_{2}+C_{3}+C_{G}$, and
$V=[v_A(k),v_B(k)]^T$ is the Bloch-like states for the potential distributions on the A and B nodes.
$\bm{a}_{1,2}$ are the basis vectors of the 2D lattice as shown in Fig.~(\ref{fig:2D_nonHermitian_CircuitLattice}).
The details of derivation of Eqs.~(\ref{eq:HV=eV}-\ref{eq:d1d2gamma}) are given in
Appendix~\ref{appendix:eq151617}.
Comparing Eq.~(\ref{eq:d1d2gamma}) with Eq.~(\ref{eq:HK_2D_model}), we
find that the parameters in these two equations can be related as
$t=-C_2$, $t_g=-(C_1+C_3/2)$ and $\gamma_y=-C_3/2$.
Therefore, by tuning capacitors $C_{1,2,3}$, one can realize the nodal points and bulk Fermi arc states in the 2D non-Hermitian electrical-circuit honeycomb lattice.
The 3D non-Hermitian honeycomb lattice to simulate the nodal drumhead with exceptional edges can be constructed by the same method as well.

\section{Conclusion\label{sec:conclusion}}
In this work we investigated possible exotic non-hermitian quantum states on 2D and 3D honeycomb lattices models with only nearest-neighbor hoppings being considered. More specifically, the bulk Fermi-arc states connecting the exceptional points, and the bulk drumhead states bounded by the exceptional lines were found in 2D and 3D cases, respectively.
By investigating the bulk-boundary correspondence of these models with open boundary conditions, we observed that the above exotic states are undermined, indicating the periodic boundary conditions are essential for the existence of these exceptional points and open nodal manifolds.  Since periodic conditions are actually unrealistic for conventional systems, such as real materials, photonic crystals and cold atoms in optical lattices,
we therefore proposed the electrical-circuit simulations, which have the advantage of easily achieving periodic boundary conditions, to realize the exotic states. Moreover, the constructed electrical circuits in principle can be easily fabricated experimentally, since all components and their usage are conventional.
%The operational amplifiers, used as voltage followers, together with capacitors and inductors consist of the
%2D and 3D non-Hermitian  honeycomb circuit lattices.
%The Kirchhoff equations are proved the same as the model Hamiltonians for the non-Hermitian honeycomb lattices models.
%We expect the exceptional points/lines states and the related bulk nodal manifolds to be realized in the proposed circuit systems.

\section*{Acknowledgments}
This work was supported by
the National Key Research and Development Program of China (No. 2017YFA0304700, No.2017YFA0303402),
the National Natural Science Foundation of China (No. 11674077, No. 11874048).
The numerical calculations in this work have been done on the supercomputing system in the Supercomputing Center of Wuhan University.

\appendix

% \section*{Supplementary Materials}

% The supplementary materials are organized as follows.
% In Sec.~\ref{sec:openBC_2D}, the details of the band structures for a strip of 2D
% non-Hermitian honeycomb lattice are given. The non-Hermitian skin effect are discussed.
% An auxiliary Hamiltonian are obtained to recover the bulk-boundary correspondence for the 2D lattice.
% In Sec.~\ref{sec:openBC_3D}, the calculated band structures for a slab of 3D
% non-Hermitian honeycomb lattice are given.
% In Sec.~\ref{sec:eq151617}, the details of the derivation of Eqs.~(15-17) in the main text are given.

\section{Edge states and skin effect in a strip of 2D non-Hermitian honeycomb lattice\label{appendix:openBC_2D}}

Considering the strip of 2D honeycomb lattice with zigzag edge in the $x$-direction, the Hamiltonian can be written as
\begin{equation}
\begin{aligned}\label{eq:2D_H_OBC}
H =\sum_{j=1}^{N} &\left(tc_{jA,k_y}^{\dagger}c_{jB,k_y}e^{-ik_ya_{1y}}+h.c.\right) \\
+\sum_{j=1}^{N-1} &\big( (t_g-\gamma_y)c_{(j+1)B,k_y}^{\dagger}c_{jA,k_y} \\
&     +(t_g+\gamma_y)c_{jA,k_y}^{\dagger}c_{(j+1)B,k_y}  \big),
\end{aligned}
\end{equation}
where $N$ is the number of unit cell in the $x$-direction.
The band dispersions for Hamiltonian~(\ref{eq:2D_H_OBC}) are shown in Fig.~\ref{fig:2D_edge_state}.
It clear that the band-crossing points (blue stars) are not correspond to the exceptional points (red dots), and the number of gap-closing point can be not equal to the number of exceptional points.
The bulk Fermi-arc states connecting exceptional points $E_1$-$E_2$ and $E_3$-$E_4$ (red dots) disappear.
But edge states arise and connect a pair of the new gap closing points (blue stars), instead of connecting the projection of the exceptional points.
These anomalies can be resolved by using the auxiliary Hamiltonian proposed in reference
\cite{WangZhong_1D_nonHermitian_PRL_2018}.
Taking a similarity transformation to $H$, we obtain
$\tilde{H}=P^{-1}HP$,
where $P$ is a $2N\times 2N$ diagonal matrix
$P=\mathrm{diag}[1,1,\alpha,\alpha,\cdots,\alpha^{N-1},\alpha^{N-1}]$ and
$\alpha=\sqrt{|(t_g-\gamma_y)/(t_g+\gamma_y)|}$.
The transformed Hamiltonian $\tilde{H}$ has explicit form as
\begin{equation}\label{eq:3D_Hermitian_H_tb_dz0}
\tilde{H}=\sum_{j=1}^{N}tc_{jA,k_y}^{\dagger}c_{jB,k_y}e^{-ik_ya_{1y}}
+\sum_{j=1}^{N-1}\tilde{t}_g c_{(j+1)B,k_y}^{\dagger}c_{jA,k_y}+h.c..
\end{equation}
After taking Fourier transform in the $x$ direction, one obtains
\begin{equation}\label{eq:2D_Hermitian_Hk}
\tilde{H}(\bm{k})=\tilde{d}_x(\bm{k})\sigma_x+\tilde{d}_y(\bm{k})\sigma_y,
\end{equation}
where
$\tilde{d}_x=\tilde{t}_g+t(\cos\bm{k}\cdot\bm{a}_1+\cos\bm{k}\cdot\bm{a}_2)$,
$\tilde{d}_y=            t(\sin\bm{k}\cdot\bm{a}_1+\sin\bm{k}\cdot\bm{a}_2)$, and
$\tilde{t}_g= \sqrt{t_g^2-\gamma_y^2}$.
Hence, the Eq.~(\ref{eq:3D_Hermitian_H_tb_dz0}) and (\ref{eq:2D_Hermitian_Hk}) is Hermitian
if $|t_g| \ge |g_y|$ is satisfied.
The gap closing points calculated from $\tilde{H}(\bm{k})$ are consistent with the
gap closing points of the strip structure as shown in Fig.~\ref{fig:2D_edge_state}.
One can calculate the Berry phase $\phi(k_y)$ for $\tilde{H}(k_x,k_y)$ with $k_y$ fixed.
$\phi(k_y)=\pi$ reveals that the edge states exist on the boundary, while $\phi(k_y)=0$
indicates no edge states existing.
Therefore, the bulk-boundary correspondence is recovered by using $\tilde{H}(\bm{k})$.

\begin{figure}[h]
	\centering
	\includegraphics[width=0.99\linewidth]{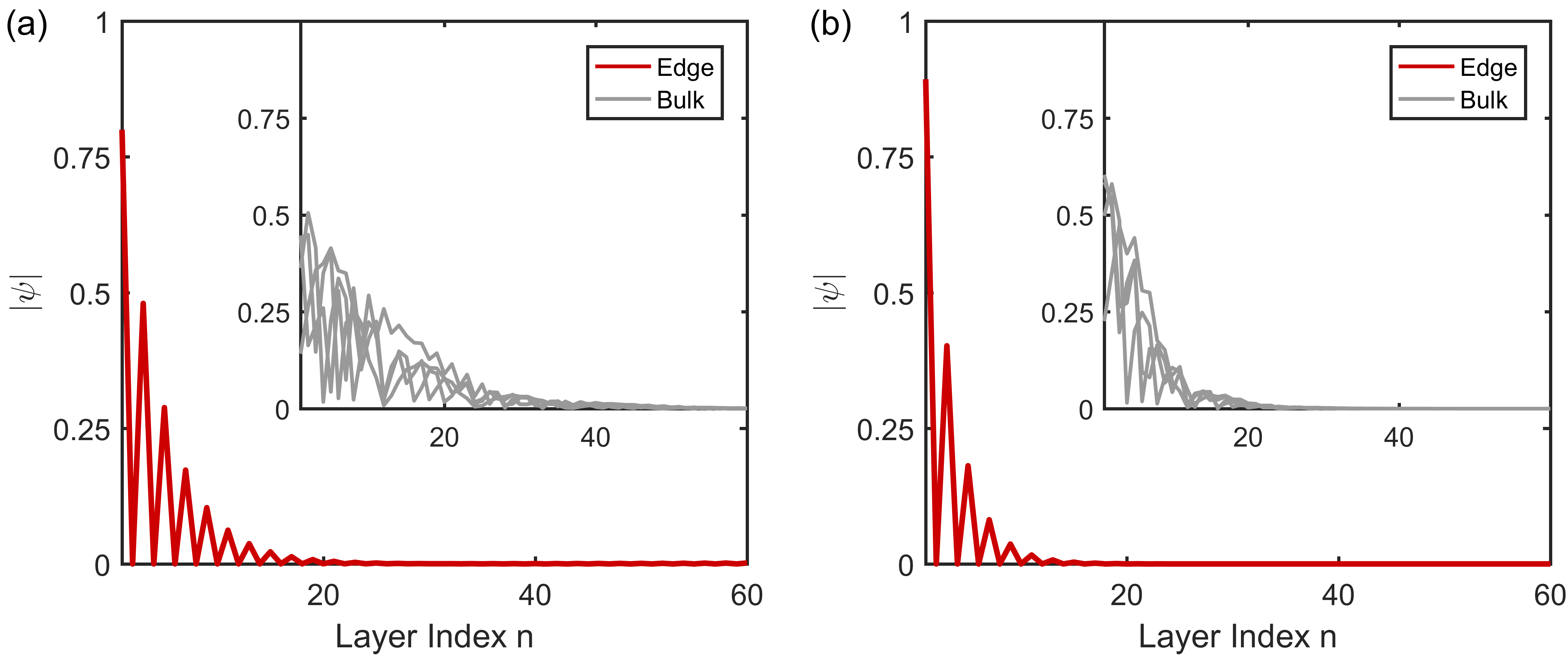}
	\caption{(a) The two edge states(red lines) is localized at the left side as expected and degenerate for the chiral symmetry with $\gamma_y=0.3$, while the bulk states (gray lines in the inset) are also localized on the boundary.
		(b) With $\gamma_y=0.6$, all the bulk and edge states are localized more heavily.}
	\label{fig:BE_G}
\end{figure}

Now, we show the skin effect for the non-Hermitian 2D honeycomb lattice with open-boundary conditions.
Considering the similarity transformation $\tilde{H}=P^{-1}HP$.
If $\tilde{H}|\psi\rangle=\lambda|\psi\rangle$, then
$P^{-1}H P|\psi\rangle=\lambda|\psi\rangle \Rightarrow H P|\psi\rangle=\lambda P|\psi\rangle$.
Thus, if $|\psi\rangle$ is an eigenvector of $\tilde{H}$ with eigenvalue $\lambda$,
then $P|\psi\rangle$ is an eigenvector of $H$ with the same eigenvalues.
With periodical boundary conditions, all states in both Hermitian and non-Hermitian 2D honeycomb lattice are Bloch waves, ensured by the translational symmetry of the lattice.
%
% From the procedure of similarity transformation, we noticed the transformation from $| \psi \rangle$ to $| \tilde{\psi}\rangle$ by matrix $P$ is not unitary, for $0<\alpha <1$.
With open boundary conditions, the bulk states of $\tilde{H}$,
$\psi=\left(\psi_{1B,k_y},\psi_{1A,k_y},\cdots,\psi_{NB,k_y},\psi_{NA,k_y}\right)^{T}$,
are still nearly Bloch-type, when the number of layers is large enough.
However, this is not the same as in non-Hermitian cases.
We can find that the wave function $P{\psi}$ of $H$ becomes
$P{\psi}_{nA,B}=\alpha^{n-1}\psi_{nA,B}$.
As $|\alpha|\ne 1$. Therefore we get that $P{\psi}$ is localized at one side of the effective 1D system (as shown in Fig.~\ref{fig:BE_G}), dubbed as ``non-Hermitian skin effect''
\cite{boundary_xiong_why_2018,WangZhong_1D_nonHermitian_PRL_2018}.
%% FIGURE5

Below we give a more intuitive way to understand the skin effect. For the non-Hermitian originated from asymmetric hopping terms $t_g\pm\gamma_y$,
the particles have larger hopping probability in a specific direction.
Although, the wave functions are Bloch type in the periodical boundary conditions, the particles accumulate to one side of the system in the open boundary conditions.
The states largely deviate from the bulk Bloch type, therefore, the breakdown of the correspondence between bulk exceptional points with periodical boundary conditions and the edge states with open boundary condition is not surprising.

%% fig---7
\begin{figure}[t]
	\centering
	\includegraphics[width=0.95\linewidth]{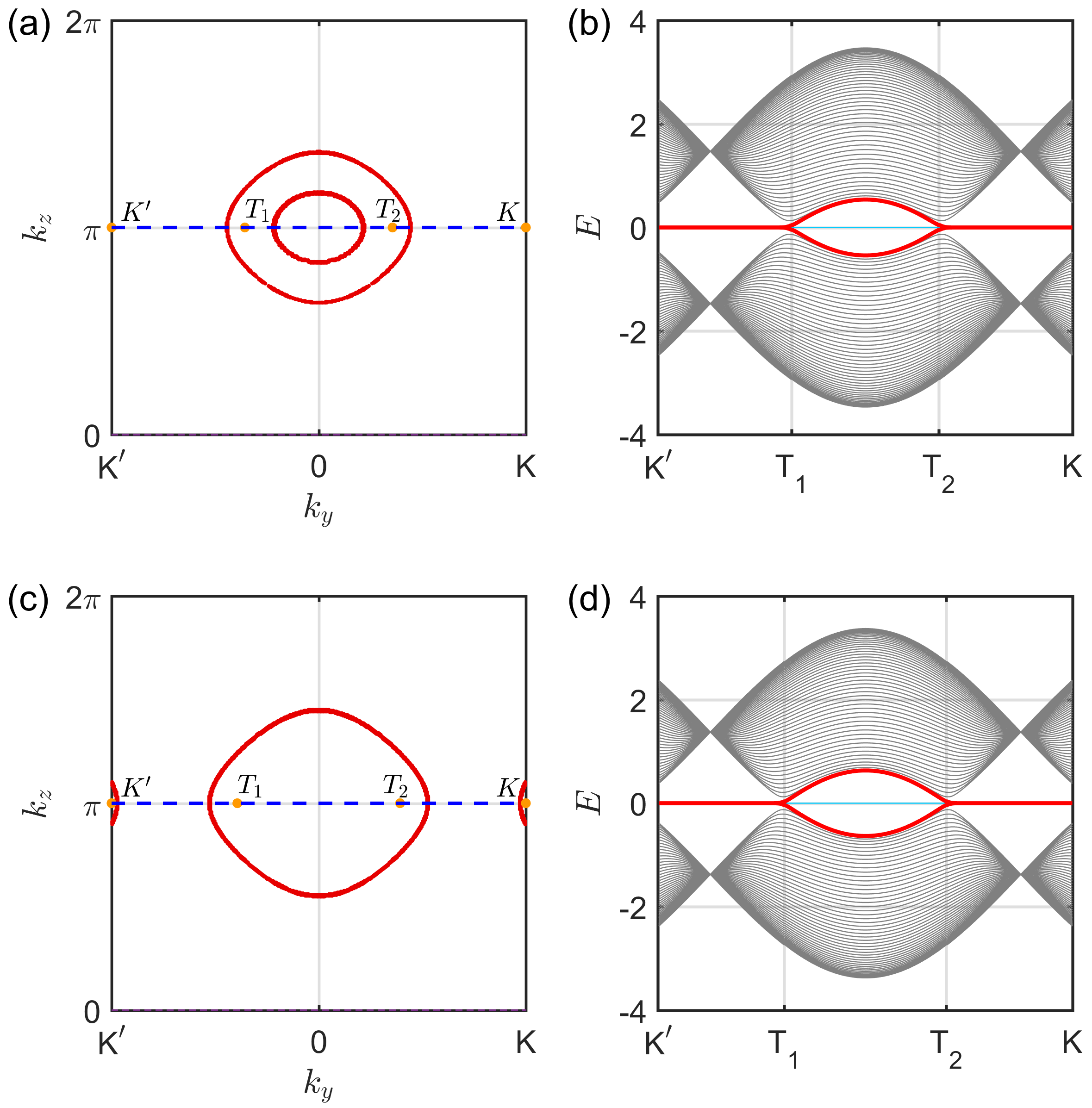}
	\caption{\label{fig:3D_surface_states_dz_0}
		For the $d_z=0$ case with $t_g=2.5$ and $t=1$, there are
		(a) two exceptional rings  for $\gamma_y=0.3$ and
		(c) one exceptional ring (red curves) for $\gamma_y=0.6$ projected to the surface BZ .
		(b, d) The slab band structures are calculated along $k_y$. The gap-closing points
		are located at $T_1$ and $T_2$ (pink points indicated in (a) and (c), not at the exceptional points.
	}
\end{figure}

%%Breakdown and Recovery of Bulk-Surface Correspondence
\section{Surface states for slab structure of 3D non-Hermitian honeycomb lattice}\label{appendix:openBC_3D}

We consider slab geometry terminated in the $x$ direction of the 3D non-Hermitian honeycomb lattice.
The band structures of the slab are calculated as shown in Fig.~\ref{fig:3D_surface_states_dz_0}.
The bulk exceptional lines (red color curves) are projected to the surface BZ as shown in Fig.~\ref{fig:3D_surface_states_dz_0}(a, c).
The gap closing points for the slab band structures are located at $T_1$ and $T_2$ points instead of at the exceptional points.
The bulk drumhead states are damaged, with no corresponding states on the surface.
While new surface states (red color bands), connecting $T_1$ and $T_2$ points, emerge as shown in Fig.~\ref{fig:3D_surface_states_dz_0}(b, d).

%%%%%%%%%%%%%%%%%%%%%%%%%%%%%%%%%%%%%%%%%%%%%%%%%%%%%%%%%%%%%%%%%%%%%%%%%%%%%%%%%%%%%%%%%%%%%%
\section{Details of the derivation of Eqs.~(15-17) in the main text}\label{appendix:eq151617}

% \begin{figure}[]
%     \centering
%     \includegraphics[width=0.99\linewidth]{fig_S3}
%     \caption{A 2D circuit lattice consist of operational amplifiers and capacitors.
%     The dashed pink box indicates the basic non-Hermitian unit cells.
%     The capacitors $C_2$ connect the unit cell and form a honeycomb-type lattice.
%     The potential distribution on the A and B nodes are denoted as $v_A(\bm R)$ and $v_B(\bm R)$, respectively.}\label{fig:S2D_nonHermitian_CircuitLattice}
% \end{figure}

The currents, which flow into nodes A and B in the cell located at $\bm{R}=0$
of the circuit lattice (shown in Fig.~\ref{fig:2D_nonHermitian_CircuitLattice}), are given as
\begin{equation}
\begin{aligned}
I_{A}(0) & =j\omega(C_{1}+C_{3})[v_{B}(0)-v_{A}(0)] \\
& +j\omega  C_{2}[v_{B}(-\bm{a}_{1})-v_{A}(0)]+j\omega C_{2}[v_{B}(-\bm{a}_{2})-v_{A}(0)] \\
& +j\omega C_{G}[0-v_{A}(0)]+\frac{1}{j\omega L}[0-v_{A}(0)],\label{eq:I_A0}
\end{aligned}
\end{equation}
\begin{equation}
\begin{aligned}
I_{B}(0) & =j\omega C_{1}[v_{A}(0)-v_{B}(0)] \\
& +j\omega  C_{2}[v_{A}(\bm{a}_{1})-v_{B}(0)]+j\omega C_{2}[v_{A}(\bm{a}_{2})-v_{B}(0)] \\
& +j\omega(C_{3}+C_{G})[0-v_{B}(0)]+\frac{1}{j\omega L}[0-v_{B}(0)],\label{eq:I_A0-1}
\end{aligned}
\end{equation}
where $\omega$ is the frequency for the sinusoidal signal, $j\equiv\sqrt{-1}$,
the vectors $\bm{a}_{1},$ \textbf{$0$}, $\bm{a}_{2}$
in the parentheses corresponding to lattice sites.
The relations for the nodes currents $I_{A,B}(\bm R)$ and the potential distributions $v_{A,B}(\bm R)$ on the whole lattice
can be obtained with the same method.

Kirchhoff's law demands that $I_{A}(\bm R)$ and $I_{B}(\bm R)$ are zero.
Therefore writing above equations into a matrix form, we get a
tight-binding-like Hamiltonian equation

\begin{widetext}
	\begin{equation}
	\left[\begin{array}{cccccccc}
	\\
	\\
	\ddots &  &  & \vdots\\
	& -C_{2} & -C_{2} & C_{s} & -(C_{1}+C_{3})\\
	&  &  & -C_{1} & C_{s} & -C_{2} & -C_{2}\\
	&  &  &  & \vdots &  &  & \ddots\\
	\\
	\\
	\end{array}\right]\left[\begin{array}{c}
	\vdots\\
	v_{B}(-\bm{a}_{1})\\
	v_{B}(-\bm{a}_{2})\\
	v_{A}(0)\\
	v_{B}(0)\\
	v_{A}(\bm{a}_{1})\\
	v_{A}(\bm{a}_{2})\\
	\vdots
	\end{array}\right]=\frac{1}{\omega^{2}L}\left[\begin{array}{c}
	\vdots\\
	v_{B}(-\bm{a}_{1})\\
	v_{B}(-\bm{a}_{2})\\
	v_{A}(0)\\
	v_{B}(0)\\
	v_{A}(\bm{a}_{1})\\
	v_{A}(\bm{a}_{2})\\
	\vdots
	\end{array}\right].\label{eq:H_tb_Circuit}
	\end{equation}
\end{widetext}
The hopping terms can be extracted from the left matrix as listed below:
$H_{AA}(\bm{R}=0)=C_{s}\equiv C_{1}+2C_{2}+C_{3}+C_{G}$,
$H_{AB}(\bm{R}=0)=-(C_{1}+C_{3})$,
$H_{AB}(\bm{R}=-{\bm{a}_{1}})=-C_{2}$,
$H_{AB}(\bm{R}=-{\bm{a}_{2}})=-C_{2}$,
$H_{BB}(\bm{R}=0)=C_{S}$,
$H_{BA}(\bm{R}=0)=-C_{1}$,
$H_{BA}(\bm{R}={\bm{a}_{1}})-C_{2}$, and
$H_{BA}(\bm{R}={\bm{a}_{2}})=-C_{2}$,
where $\bm{R}$ is the lattice vector and $H_{nm}(\bm{R})$
are the tight-binding parameters between node $n$ located at the
home unit cell and node $m$ located at $\bm{R}$. With these
terms, the Hamiltonian in the $\bm{k}$ space can be
obtained by the Fourier transform $H_{nm}(\bm{k})=\sum_{\bm{R}}e^{i\bm{k}\cdot\bm{R}}H_{nm}(\bm{R})$, leading to
\begin{equation}
\begin{aligned}
H_{AA}(\bm{k})&=H_{BB}(\bm{k})=C_{s},\\
H_{AB}(\bm{k})&=-(C_{1}+C_{3})-C_{2}(e^{-i\bm{k}\cdot{{\bm{a}_{1}}}}+e^{-i\bm{k}\cdot\bm{a}_{2}}),\\
H_{BA}(\bm{k})&=-C_{1}-C_{2}( e^{i\bm{k}\cdot\bm{a}_{1}}+e^{i\bm{k}\cdot\bm{a}_{2}} )
.\label{eq:HBAk}
\end{aligned}
\end{equation}
Rewriting the matrix in terms of the Pauli matrices, we
obtain
\begin{equation}
H(\bm{k})=C_s\sigma_0+d_{x}(\bm{k})\sigma_{x}+(d_{y}(\bm k)+i\gamma_{y})\sigma_{y},\label{eq:Hk_sigma}
\end{equation}
where
\begin{equation}
\begin{aligned}
d_{x}(\bm{k})&=-(C_{1}+\frac{C_{3}}{2})-C_{2}(\cos\bm{k}\cdot\bm{a}_{1}+\cos\bm{k}\cdot\bm{a}_{2}),\\
d_{y}(\bm{k})&=-C_{2}(\sin\bm{k}\cdot\bm{a}_{1}+\sin\bm{k}\cdot\bm{a}_{2}),\\
\gamma_{y}&=-\frac{C_{3}}{2}.\label{eq:d1d2}
\end{aligned}
\end{equation}
Comparing Eq.~(\ref{eq:d1d2}) with Eq.~(1), we get $t=-C_{2}$, $t_g=-(C_{1}+\frac{C_{3}}{2})$, and
$\gamma_{y}=-\frac{C_{3}}{2}$ as given in the main text.

% \end{document}

%%%%%%%%%%%%%%%{}
\bibliography{refs}
\bibliographystyle{apsrev4-1}

\end{document}